\documentclass[aps,reprint,superscriptaddress,pra]{revtex4-2}

\usepackage[utf8]{inputenc}
\usepackage{amsmath}
\usepackage{amsfonts}
\usepackage{amssymb}
\usepackage{bm}
\usepackage{color}
\usepackage{tcolorbox}
\usepackage{graphicx}
\usepackage{soul}
\usepackage{CJK}
\usepackage{physics}
\usepackage{mathrsfs}
\usepackage{multirow}
\usepackage{xcolor}
\usepackage{mathtools}
\usepackage{lipsum}
\usepackage{comment}
\usepackage{booktabs}
\usepackage{textcomp}

\usepackage[colorlinks,citecolor=blue,linkcolor=red,anchorcolor=blue,urlcolor=blue]{hyperref}

\begin{document}
\begin{CJK*}{UTF8}{gbsn}
\title{Berry-Phase-Induced Chirality in Thermodynamics}
        
\author{Zhaoyu Fei (费兆宇)}
\affiliation{Zhejiang Key Laboratory of Quantum State Control and Optical Field Manipulation, Department of Physics, Zhejiang Sci-Tech University,
310018, Hangzhou, People’s Republic of China}

\author{Yu-Han Ma (马宇翰)}
\email{yhma@bnu.edu.cn}
\affiliation{School of Physics and Astronomy, Beijing Normal University, Beijing, 100875, China}
\affiliation{Key Laboratory of Multiscale Spin Physics (Ministry of Education), Beijing Normal University, Beijing 100875, China}

\begin{abstract} 
Geometric phases are foundational to isolated quantum systems, yet their thermodynamic role in open systems remains unrevealed.
Developing a dissipative adiabatic perturbation expansion, we discover a Berry-phase-induced chiral work difference that survives decoherence. This chirality evolves from an interferometric thermodynamic Aharonov-Bohm effect in the unitary regime to a fringe-free signal in the dissipative regime. We illustrate this framework in a two-level system and assess its experimental feasibility. Our findings clarify the role of quantum geometry in the geometric formulation of thermodynamics.
\end{abstract}

\maketitle

\textit{Introduction}.---The geometric formulation of physical laws represents one of the most profound unifying principles in modern physics. A quintessential manifestation of this geometrization in the quantum realm is the Berry phase~\cite{Berry1984}. As an intrinsic geometric property of the Hilbert space, it reveals that a quantum state does not merely accumulate a dynamical phase during its adiabatic evolution; rather, it picks up a geometric phase governed by the curvature of the parameter manifold~\cite{Resta1994,Xiao2010}. This phenomenon is exemplified in the Aharonov-Bohm effect~\cite{Aharonov1959}, where the topological structure of the vector potential manifests as observable interference fringes, fundamentally altering our understanding of the relationship between potentials and fields~\cite{Peshkin1989}.

Equally profound is the geometrization of thermodynamics, which has attracted sustained interest over the past half-century~\cite{weinholdMetricGeometryEquilibrium1975,ruppeinerThermodynamicsRiemannianGeometric1979b}, spanning from equilibrium fluctuations to non-equilibrium processes~\cite{georgeRiemannianGeometryThermodynamic1995,wang2026thermodynamic}. Finite-time thermodynamic processes~\cite{Andresen1984,Bjarne2022,dann2020quantum, qiu20roadmap}, such as non-equilibrium driving or relaxation, are characterized by thermodynamic metrics, with the dissipated work or heat bounded by the geometric distance on the control-parameter manifold~\cite{Salamon1984, Crooks2007, Sivak2012,Naoto2019,shiraishi2023introduction}. Recently, bridging the geometric formulations of quantum mechanics and thermodynamics has garnered increasing attention~\cite{Scandi2019, Abiuso2020e, 2022AnzaPRE}, with a particular focus on its implications for quantum thermal machines~\cite{Ren2010,Abiuso2020, Brandner2020, Bhandari2020}. 

Despite these advances, the emergence of dissipation across the quantum-to-classical crossover remains a fundamental open question. As Berry demonstrated~\cite{Berry1993}, adiabatic driving in quantum systems generates a first-order geometric magnetism that acts orthogonally to the driving velocity, thereby yielding zero work. Crucially, this geometric force remains non-dissipative even when the system is coupled to a thermal bath~\cite{Campisi2012}. Genuine first-order classical friction, however, originates from a physically distinct contribution captured by the thermodynamic metric~\cite{Crooks2007,Scandi2019,maMinimalEnergyCost2022}. 
Given that the geometric phase manifests as a non-dissipative force, and its conventional signature of quantum interference is fragile against decoherence~\cite{Zurek2003}, one would naturally speculate its disappearance in the classical limit, leaving no observable trace in the irreversible dissipation.

In this Letter, we overturn this intuitive speculation. 
The key insight is that although the first-order geometric magnetism does no work, the Berry phase leaves its imprint at the next order. 
By developing a dissipative adiabatic perturbation expansion (DAPE) for a periodically driven open quantum system, we show that this imprint survives as a \textit{chiral work difference} and persists from unitary dynamics to strong decoherence. 
Subtracting the work of clockwise and counterclockwise protocols isolates this chirality.
In the unitary regime, this chirality manifests as an interferometric \textit{Thermodynamic Aharonov-Bohm Effect}: the chiral work difference oscillates with the driving period, producing fringes governed by the dynamical and geometric phases. In the dissipative regime, while decoherence washes out these fringes, the chiral work difference robustly persists in the non-equilibrium periodic steady state, enabling a purely thermodynamic readout of the Berry phase.

\textit{Dissipative Adiabatic Perturbation Expansion.} ---Considering on open quantum system driven by a set of time-dependent control parameters $\mathbf{R}(t)=[R^{1}(t),R^{2}(t),...,R^{i}(t),...]$, with its instantaneous Hamiltonian $\hat{H}(\mathbf{R})=\sum_n E_n\ket{n(\mathbf{R})}\bra{n(\mathbf{R})}$. Here, the eigenenergies $E_n$ are are assumed to be constant and non-degenerate. Since the level spacings $\omega_{mn}=(E_m-E_n)/\hbar$ are fixed along the protocol, the work supplied by the external agent stems solely from the geometric rotation of the instantaneous eigenbasis $\{\ket{n(\mathbf{R})}\}$. Characterizing this state evolution via $\partial_t\ket{n}=-\mathrm{i}\sum_m A_{mn}\ket{m}$, where $A_{mn}=\mathrm{i}\braket{m}{\partial_t n}$ is the Berry connection matrix elements and $\partial_t\equiv\partial/\partial t$, the average work $W(t)\equiv\int_0^t \operatorname{Tr}[\hat{\rho}(t')\,\partial_{t'}\hat{H}(t')]\mathrm{d}t'$~\cite{Alicki1979,su2018heat,Seegebrecht2024} performed on the system with density operator $\hat{\rho}(t)$ is
\begin{equation}
W(t)=-\hbar\sum_{m\neq n}\omega_{mn}\int_0^t \operatorname{Im}[A_{nm}(t')\,\rho_{mn}(t')]\mathrm{d}t'.
\label{eq:Woff_def}  
\end{equation}
Here, the matrix elements $\rho_{mn}(t)\equiv\bra{m(t)}\hat{\rho}(t)\ket{n(t)}$ obey a Born–Markov master equation when the system is weakly coupled to a thermal bath at inverse temperature $\beta$. In the instantaneous eigenbasis, ignoring the Lamb shift, the diagonal and off‑diagonal ($m\neq n$) elements evolve as~\cite{breuer2002theory, Albash2012, Yamaguchi2017}
\begin{gather}
  \begin{split}
\dot\rho_{nn}(t) =& \sum_m[w_{nm}\rho_{mm}(t)-w_{mn}\rho_{nn}(t)] \\
               &+\mathrm{i}\sum_{k\neq n}[A_{nk}(t)\rho_{kn}(t)-\rho_{nk}(t)A_{kn}(t)], \\[4pt]
\dot\rho_{mn}(t) =& -z_{mn}(t)\,\rho_{mn}(t)+\mathrm{i}A_{mn}(t)[\rho_{nn}(t)-\rho_{mm}(t)] \\
               &+ \mathrm{i}\sum_{k\neq m,n}[A_{mk}(t)\rho_{kn}(t)-\rho_{mk}(t)A_{kn}(t)],
\label{eq:offdiag}
    \end{split}
\end{gather}
where $w_{mn}$ are the population relaxation rates,
$\gamma_{mn}$ the dephasing rates, 
$z_{mn}(t)=\mathrm{i}[\omega_{mn}-A_{mm}(t)+A_{nn}(t)]+\gamma_{mn}$ the complex frequency, and $\omega_{mn}\gg\gamma_{mn}$ is required. For simplicity, $w_{mn}$ and $\gamma_{mn}$ are taken to be time‑independent. 
Initially the system is prepared in the instantaneous thermal state:
$\rho_{mn}(0)=0\;(m\neq n)$,
$\rho_{nn}(0)=\mathrm{e}^{-\beta E_n}/\sum_n\mathrm{e}^{-\beta E_n}$.

We focus on the slow‑driving regime, characterized by the small adiabaticity parameter $\epsilon \equiv 2\pi/(\omega T) \ll 1$, where $\omega$ is the typical level spacing, and $T$ is the driving period of $\mathbf{R}(t)$. Within the DAPE framework we expand the off‑diagonal coherences as
\begin{equation}
\rho_{mn}(t) = \rho_{mn}^{(\rm p)}(t) + \rho_{mn}^{(\rm f)}(t) + \mathcal{O}(\epsilon^3),
\label{eq:expansion}
\end{equation}
where $\rho^{(\rm p)}_{mn}$ is driven by the population‑difference term
$\mathrm{i}A_{mn}(t)[\rho_{nn}(t)-\rho_{mm}(t)]$ in Eq.~\eqref{eq:offdiag},
and $\rho_{mn}^{(\rm f)}$ encodes the multilevel feedback from
$\mathrm{i}\sum_{k\neq m,n}[A_{mk}(t)\rho_{kn}(t)-\rho_{mk}(t)A_{kn}(t)]$. Because the relaxation rates satisfy detailed balance, the populations evolve slowly with nonadiabatic corrections $\rho_{mm}(t)-\rho_{mm}(0)\sim\mathcal{O}(\epsilon^2)$~\cite{Sun1988, Rigolin2008}.
To leading order we may therefore replace the time‑dependent population difference by the constant initial value,
$\rho_{mm}(t)-\rho_{nn}(t)\simeq\rho_{mm}(0)-\rho_{nn}(0)\equiv\Delta_{mn}$.

Substituting the expansion \eqref{eq:expansion} for $\rho_{mn}$ into Eq.~\eqref{eq:Woff_def} decomposes the average work into a sum of memory integrals, $W=W^{(\rm p)}+W^{(\rm f)}$. Integrating by parts eliminates the memory integrals order by order and reduces the average work to local boundary terms, as detailed in \textit{Appendix A} of the \textbf{End Matter}. For a closed parameter loop $\mathbf{R}(T)=\mathbf{R}(0)$, we obtain the compact result $W^{(\rm p)}=W^{(\rm pm)}+W^{(\rm pb)}+\mathcal{O}(\epsilon^3)$ with
\begin{gather}
  \begin{split}
W^{(\rm pm)}=&
\sum_{m\neq n} \left[\cos \varphi_{mn}Q_{mn}+
\gamma_{mn}\!\int_0^T\!\!Q_{mn}(t)\mathrm{d}t\right],\\
W^{(\rm pb)}=&
\sum_{m\neq n} \Bigg[-Q_{mn}\mathrm{e}^{-\gamma_{mn}T}\cos\Theta_{mn}
+\int_0^T\Lambda_{mn}(t)\mathrm{d}t\Bigg].
\label{eq:W1_closed}
    \end{split}
\end{gather}
Here, we have introduced the abbreviations:
$\Theta_{mn}\equiv \omega_{mn}T-\varphi_{mn}-\Phi_{mn}$,
$\varphi_{mn}\equiv \arctan[2\omega_{mn}\gamma_{mn}/(\omega_{mn}^2-\gamma_{mn}^2)]$ (the ``dissipative angle'' in $\Theta_{mn}$),
$Q_{mn}(t)\equiv \hbar\omega_{mn}\Delta_{mn}|A_{mn}(t)|^2/(\omega_{mn}^2+\gamma_{mn}^2)$,
$Q_{mn}\equiv Q_{mn}(0)$,
and
$\Lambda_{mn}(t)\equiv \sin\varphi_{mn}Q_{mn}(t)[A_{mm}(t)-A_{nn}(t)+\partial_{t}\arg A_{mn}(t)]$. And the Berry phase difference for the closed loop is
\begin{gather}
  \begin{split}
\Phi_{mn}\equiv& \int_0^T [A_{mm}(t)-A_{nn}(t)]\mathrm{d}t\\
=& i\oint [ \bra{m}\nabla_{\mathbf{R}}\ket{m} - \bra{n}\nabla_{\mathbf{R}}\ket{n}] \cdot d\mathbf{R}.
\label{eq:Phi_def}
    \end{split}
\end{gather}
Equation~\eqref{eq:W1_closed} clearly separates the average work into two distinct geometric origins: a metric contribution $W^{(\rm pm)}$ and a Berry-phase contribution $W^{(\rm pb)}$. 

In Eq.~\eqref{eq:W1_closed}, $Q_{mn} \propto |A_{mn}|^2 =\sum_{i,j} g_{ij}^{mn} \dot{R}^i \dot{R}^j \sim \mathcal{O}(T^{-2})$, where $g_{ij}^{mn} \equiv \Re[\bra{m}\partial_i\ket{n}\bra{n}\partial_j\ket{m}]$ ($\partial_i\equiv \nabla_{R_i}$). Thus, the leading-order $\mathcal{O}(\epsilon)$ dissipated work is governed solely by the time integral. Using $\gamma_{mn} Q_{mn}= \hbar \Delta_{mn} \sin\varphi_{mn} |A_{mn}|^2/2$, this integral reduces to
\begin{equation}
W^{(\rm pm)} = \hbar\int_0^T\dot{s}^2\mathrm{d}t+\mathcal{O}(\epsilon^2),
\end{equation}
where $\dot{s}$ is the thermodynamic speed defined by the Riemannian line element $\mathrm{d}s^2= \sum_{i,j}\mathcal{G}_{ij}\mathrm{d}R^i \mathrm{d}R^j$, introducing the thermodynamic metric on the control parameter manifold as
\begin{equation}
\mathcal{G}_{ij}\equiv \sum_{m< n}\Delta_{mn}\sin\varphi_{mn} g_{ij}^{mn}.
\label{metric}
\end{equation}
Applying the Cauchy-Schwarz inequality $\int_0^T \dot{s}^2 \mathrm{d}t \ge T^{-1}(\int_0^T \dot{s}\,\mathrm{d}t)^2$ directly yields
\begin{equation}
W^{(\rm pm)}\geq \hbar\frac{ \mathcal{L}^2}{T}+\mathcal{O}(\epsilon^2), \quad \text{where} \quad \mathcal{L} \equiv \oint \mathrm{d}s. \label{eq:wpmg}
\end{equation}
This indicates that the $\mathcal{O}(\epsilon)$ dissipation is strictly bounded by the geometric thermodynamic length $\mathcal{L}$~\cite{Salamon1984,Crooks2007}, naturally recovering the typical $1/T$-scaling irreversibility in finite-time thermodynamics~\citep{Salamon1984,maUniversalConstraintEfficiency2018,ma2020experimental,yuanOptimizingThermodynamicCycles2022,zhaoEngineeringRatchetbasedParticle2024}. This lower bound is saturated if and only if the protocol maintains a constant thermodynamic speed ($\dot{s} = \text{const}$)~\cite{Salamon1984, Sivak2012,liGeodesicPathMinimal2022}.

While $W^{(\rm pm)}$ dictates the symmetric first-order dissipation, the Berry-phase contribution $W^{(\rm pb)}$ induces a second-order $\mathcal{O}(\epsilon^2)$ chiral asymmetry. 
Reversing the protocol ($\Phi_{mn}\to -\Phi_{mn}$, $A_{mn}\to -A_{mn}$, and $\partial_t\arg A_{mn}\to -\partial_t\arg A_{mn}$) maps $W^{(\rm p)}_{\rm cw}$ to $W^{(\rm p)}_{\rm cc}$. 
Since $W^{(\rm pm)}$ is invariant under this reversal, the difference between clockwise and counterclockwise work cleanly eliminates the symmetric friction, isolating the chiral work difference
\begin{equation}
\begin{split}
\Delta W &\equiv W^{(\rm p)}_{\rm cw}-W^{(\rm p)}_{\rm cc} \\
&= 2\sum_{m\neq n} \left[ -Q_{mn}\sin\Phi_{mn}f_{mn}(T) + \int_0^T\!\!\Lambda_{mn}(t)\mathrm{d}t \right],
\end{split}
\label{eq:Delta_W_timedep}
\end{equation}
where $f_{mn}(T)\equiv \mathrm{e}^{-\gamma_{mn}T} \sin(\omega_{mn}T-\varphi_{mn})$ is a damped oscillatory envelope. For driving generated by a time-independent Hermitian operator $\hat{G}$, i.e., $\hat{H}(t)=\mathrm{e}^{\mathrm{i}\hat{G} t}\hat{H}(0)\mathrm{e}^{-\mathrm{i}\hat{G} t}$, the connection $A_{mn}$ and thermodynamic speed $\dot{s}$ are constant. Consequently, the bound in Eq.~\eqref{eq:wpmg} strictly saturates, and Eq.~\eqref{eq:Delta_W_timedep} simplifies to
\begin{equation}
\Delta W = 2\sum_{m\neq n} Q_{mn} [-\sin\Phi_{mn}f_{mn}(T) + \Phi_{mn}\sin\varphi_{mn}],
\label{eq:Delta_W_derived}
\end{equation}
directly linking the chiral response to the accumulated Berry phase. 

The influence of dephasing on this chiral signature is revealed in two opposite limits: i) In the unitary regime ($\gamma_{mn}\ll T^{-1}$), $f_{mn}(T)\simeq\sin(\omega_{mn}T)$, yielding
\begin{equation}
\Delta W^{(\rm{ u})} \simeq -2\sum_{m\neq n} Q_{mn}\sin\Phi_{mn}\sin(\omega_{mn}T).
\label{eq:Delta_weak}
\end{equation}
The chiral work difference thus encodes the interference of multi-channel non-adiabatic transitions, with each channel accumulating a dynamical phase $\omega_{mn}T$ and a geometric phase $\Phi_{mn}$, manifesting the $1/T^2$-scaling of the oscillatory extra work in finite-time quantum adiabatic processes~\cite{Chen2019,fei2022efficiency}. This interference mechanism constitutes a thermodynamic Aharonov-Bohm effect, transforming the extra work into an interferometric probe of the underlying quantum geometric phases. ii) In the dissipative regime ($\gamma_{mn}\gg T^{-1}$), coherent oscillations are exponentially suppressed ($f_{mn}(T)\simeq0$), reducing the chiral work difference to
\begin{equation}
\Delta W^{(\rm{d})} \simeq 2\sum_{m\neq n} Q_{mn}\Phi_{mn}\sin\varphi_{mn}.
\label{eq:Delta_strong}
\end{equation}
This surviving term originates from off-diagonal coherence persistently sustained by the slow driving, with the dissipative angle $\varphi_{mn}$ dictating the relative weight of each channel. 
Crucially, this overturns the intuition that the Berry phase merely generates a non‑dissipative geometric magnetism at first order. 
Instead, it ultimately governs the chirality in the dissipated work---rather than having its fringes erased by decoherence, it imprints a robust, fringe‑free signal onto the irreversible dissipation.

These two limits reveal a scaling crossover of the symmetric dissipation between the unitary ($\gamma_{mn}\to 0$) and dissipative ($\gamma_{mn}\gg T^{-1}$) regimes, where it changes from $1/T^2$ to $1/T$.
By contrast, the Berry-phase-induced chiral signature remains pinned at $\mathcal{O}(T^{-2})$ across the entire crossover.
This persistent geometric signal establishes thermodynamic chirality as a robust probe of quantum geometry. 
This persistent geometric signal establishes thermodynamic chirality as a robust probe of quantum geometry. Finally, we stress that the above derivations omit the multilevel feedback term $W^{(\rm f)}$. This $\mathcal{O}(\epsilon^2)$ term requires at least three distinct energy levels with non-vanishing cyclic products ($A_{mk}A_{kn}A_{nm} \neq 0$); it vanishes for the two-level system and can be systematically eliminated in generalized setups, see \textit{Appendix A} of the \textbf{End Matter} for details.

\textit{Application to a Driven Two-Level System.} ---Consider a Two-Level System driven by a uniformly precessing magnetic field. 
The Hamiltonian reads~\cite{su2018heat,Scandi2019,liu2020fluctuation}
\begin{equation}
   \hat H(t) = -\frac{\hbar \omega}{2}  \mathbf{n}(t) \cdot \hat{\boldsymbol{\sigma}} \label{eq:hamiltonian_tls}
\end{equation}
where $\mathbf{n}(t) = (\sin\theta \cos\phi, \sin\theta \sin\phi, \cos\theta)$ is the unit vector precessing with azimuthal angle $\phi(t) = \Omega t$ ($\Omega = 2\pi/T$) at a constant polar angle $\theta$, and $\hat{\boldsymbol{\sigma}}=(\hat\sigma_x, \hat\sigma_y, \hat\sigma_z)$ is the Pauli matrix vector. The corresponding instantaneous ground and excited eigenstates are defined as $\ket{+} = \big(\cos\frac{\theta}{2}, \mathrm{e}^{\mathrm{i}\phi}\sin\frac{\theta}{2}\big)^\mathsf{T}$ and $\ket{-} = \big(\sin\frac{\theta}{2}, -\mathrm{e}^{\mathrm{i}\phi}\cos\frac{\theta}{2}\big)^\mathsf{T}$, separated by a constant energy gap $\hbar\omega$. Because the driving is generated by the constant operator $\hat{G} = \Omega \sigma_z/2$, the off-diagonal Berry connections are time-independent, allowing a direct implementation of Eq.~\eqref{eq:Delta_W_derived}. 

The geometry of the parameter manifold spanned by $(\theta, \phi)$ seamlessly dictates the macroscopic dissipation. Based on the adiabatic eigenstates, the metric component evaluates to $g_{\phi\phi} = |\langle - | \partial_\phi | + \rangle|^2 = \sin^2\theta/4$. With $\Delta = \tanh(\beta\hbar\omega/2)$, $\gamma=\gamma_{+-}=\gamma_{-+}$ denoting the dephasing rate (taken constant, assuming isotropic system‑bath coupling and hence an orientation‑independent thermalization), and 
$\varphi = \arctan[2\omega\gamma/(\omega^2-\gamma^2)]$
Eq.~\eqref{metric} yields the thermodynamic length over the closed loop
\begin{equation}
\mathcal{L} = \oint \sqrt{ \Delta \sin\varphi \, g_{\phi\phi}} \, \mathrm{d}\phi = \pi\sin\theta\sqrt{\frac{2\Delta\omega\gamma}{\omega^2+\gamma^2}}.
\label{LTSL}
\end{equation}
The coefficient $Q = Q_{+-}=Q_{-+}$ reads $Q = \hbar\mathcal{L}^2 / (2\gamma T^2)$. Substituting this into Eq.~\eqref{eq:Delta_W_derived}, the chiral work difference explicitly evaluates to
\begin{equation}
\Delta W_{\rm TLS} =\frac{2\hbar\mathcal{L}^2}{\gamma T^2} [-\mathrm{e}^{-\gamma T} \sin\Phi\sin(\omega T-\varphi) + \Phi\sin\varphi ] ,
\label{eq:Delta_W_TLS_exact}
\end{equation}
where $\Phi\equiv \oint (\bra{+}\partial_t\ket{+}-\bra{-}\partial_t\ket{-}) \mathrm{d}t= 2\pi\cos\theta$ is the Berry phase difference, which is related to the solid angle of the loop traced by the magnetic field on the Bloch sphere. 

The dependence on $\theta$ reveals distinct geometric regimes. 
For $\theta = 0$ or $\pi$, the magnetic field is static; $\mathcal{L} \propto \sin\theta = 0$ trivially yields $\Delta W_{\rm TLS} = 0$. 
At $\theta = \pi/2$, the magnetic field precesses in the equatorial plane, where $\Phi = 2\pi\cos(\pi/2) = 0$, again giving $\Delta W_{\rm TLS} = 0$. 
This cancellation is non‑trivial and originates from a chiral symmetry that relates the two driving protocols: $\sigma_x H_{\rm cw}(t) \sigma_x = H_{\rm ccw}(t)$, enforcing identical work in both directions. 
For any $\theta \neq \pi/2$, this symmetry is broken and the chiral work difference becomes finite.
These analytical findings are corroborated by an exact solution of the TLS Bloch equations, which we detail in \textit{Appendix B} of the \textbf{End Matter}.

In the unitary regime ($\gamma \ll T^{-1}$), environmental decoherence is negligible, resulting in $\mathrm{e}^{-\gamma T} \approx 1$ and a vanishing dissipative angle $\varphi$. The chiral work difference simplifies to
\begin{equation}
\Delta W_{\rm TLS}^{(\rm u)} \simeq - \frac{4\pi^2\hbar\Delta\sin^2\theta}{\omega T^2} \sin(\omega T)\sin\Phi.
\label{eq:TLS_weak}
\end{equation}
Equation~\eqref{eq:TLS_weak} captures the thermodynamic Aharonov-Bohm effect: the chiral work difference oscillates with $\omega T$, directly reflecting interference between the two transition channels. Notably, in the absence of dissipation ($\gamma=0$), our result is in agreement with the exact unitary solution reported in Ref.~\cite{su2018heat}. In the dissipative regime ($\gamma\gg T^{-1}$), the system is irreversibly slaved to the non-equilibrium periodic steady state, forcing the oscillatory envelope to vanish. The chiral work difference then reduces to
\begin{equation}
\Delta W_{\rm TLS}^{(\rm d)} \simeq \frac{8\pi^2 \hbar\omega^2\gamma\Delta\sin^2\theta}{(\omega^2+\gamma^2)^2T^2}\Phi,
\label{eq:TLS_strong}
\end{equation}
which explicitly demonstrates that, while the symmetric finite-time dissipation scales as $\mathcal{L}^2/T$, the chiral work difference emerges as a $\mathcal{O}(T^{-2})$ correction governed by $\Phi$. This realizes a fringe‑free thermodynamic readout of the geometric phase, fully consistent with the general prediction of Eq.~\eqref{eq:Delta_strong}.

To assess experimental feasibility, we estimate the chiral work difference using typical liquid-state NMR parameters (e.g.,  ${}^{13}\rm C$ nuclear spin)~\cite{2014NMR,2023NMR}. 
We take an effective energy gap $\omega/2\pi \sim 1$ kHz and a decoherence rate $\gamma/2\pi \sim 10^{-3}$ kHz. 
Despite the room-temperature environment, pseudo-thermal states with an effective spin temperature $T_{\rm eff} \sim 100$ nK can be prepared, yielding a thermal bias $\Delta \sim \mathcal{O}(10^{-1})$.
For a precession angle $\theta = \pi/4$, the thermodynamic length $\mathcal{L} \sim \mathcal{O}(10^{-2})$. 
In the unitary regime ($T \ll \gamma^{-1}$), a fast driving period $T = 10$ ms yields a single-spin chiral work $\Delta W_{\rm TLS}^{(\rm u)} /h \sim 1$ Hz; an ensemble of just $10^3$ spins amplifies this signal into the kHz regime, well within experimental resolution~\cite{2014NMR}. 
In the dissipative regime, a prolonged period $T = 10$ s gives $\Delta W_{\rm TLS}^{(\rm d)} /h \sim 10^{-7}$ Hz. 
Although this sub-$\mu$Hz single-spin signal is extremely faint, a typical macroscopic NMR ensemble of $\sim 10^{20}$ molecules cumulatively amplifies it to a robustly detectable level, confirming the viability of reading out the geometric phase on current NMR platforms.

\textit{Summary and Outlook.}---Using the DAPE framework, we have demonstrated that the Berry phase imprints a definite chiral signature on the thermodynamics of driven open quantum systems. In the unitary regime, this manifests as a thermodynamic Aharonov-Bohm effect achieved by inter-channel quantum interference. Meanwhile, decoherence enables a fringe-free readout of the Berry phase, highlighting the thermal bath as an indispensable agent that encodes quantum geometry into measurable work.
 
Looking forward, the DAPE framework opens a direct route to Berry‑phase‑induced chirality in interacting many‑body systems. 
Collective phenomena, such as high state degeneracy and collective coherence, could amplify the chiral work difference to a macroscopic quantum effect. Its experimental feasibility should therefore extend to platforms where collective amplification is intrinsic, such as interacting spin ensembles~\cite{ULYANOV1992179,ma2017quantum,ma2017quantum1} and superconducting quantum circuits~\cite{yi2026third}.

\textit{Acknowledgment}.--- Z. F. acknowledges the Science Challenge Project (No. TZ2025017), the National Natural Science Foundation of China (Grant No. 12405046), and the Science Foundation of Zhejiang Sci-Tech University (Grants No. 23062181-Y). Y.-H. M. thanks the National Natural Science Foundation of China for support under Grant No. 12305037.
\end{CJK*}
\bibliography{refs.bib}

\begin{thebibliography}{51}%
\makeatletter
\providecommand \@ifxundefined [1]{%
 \@ifx{#1\undefined}
}%
\providecommand \@ifnum [1]{%
 \ifnum #1\expandafter \@firstoftwo
 \else \expandafter \@secondoftwo
 \fi
}%
\providecommand \@ifx [1]{%
 \ifx #1\expandafter \@firstoftwo
 \else \expandafter \@secondoftwo
 \fi
}%
\providecommand \natexlab [1]{#1}%
\providecommand \enquote  [1]{``#1''}%
\providecommand \bibnamefont  [1]{#1}%
\providecommand \bibfnamefont [1]{#1}%
\providecommand \citenamefont [1]{#1}%
\providecommand \href@noop [0]{\@secondoftwo}%
\providecommand \href [0]{\begingroup \@sanitize@url \@href}%
\providecommand \@href[1]{\@@startlink{#1}\@@href}%
\providecommand \@@href[1]{\endgroup#1\@@endlink}%
\providecommand \@sanitize@url [0]{\catcode `\\12\catcode `\$12\catcode `\&12\catcode `\#12\catcode `\^12\catcode `\_12\catcode `\%12\relax}%
\providecommand \@@startlink[1]{}%
\providecommand \@@endlink[0]{}%
\providecommand \url  [0]{\begingroup\@sanitize@url \@url }%
\providecommand \@url [1]{\endgroup\@href {#1}{\urlprefix }}%
\providecommand \urlprefix  [0]{URL }%
\providecommand \Eprint [0]{\href }%
\providecommand \doibase [0]{https://doi.org/}%
\providecommand \selectlanguage [0]{\@gobble}%
\providecommand \bibinfo  [0]{\@secondoftwo}%
\providecommand \bibfield  [0]{\@secondoftwo}%
\providecommand \translation [1]{[#1]}%
\providecommand \BibitemOpen [0]{}%
\providecommand \bibitemStop [0]{}%
\providecommand \bibitemNoStop [0]{.\EOS\space}%
\providecommand \EOS [0]{\spacefactor3000\relax}%
\providecommand \BibitemShut  [1]{\csname bibitem#1\endcsname}%
\let\auto@bib@innerbib\@empty
\bibitem [{\citenamefont {Berry}(1984)}]{Berry1984}%
  \BibitemOpen
  \bibfield  {author} {\bibinfo {author} {\bibfnamefont {M.~V.}\ \bibnamefont {Berry}},\ }\bibfield  {title} {\bibinfo {title} {Quantal phase factors accompanying adiabatic changes},\ }\href {https://doi.org/10.1098/rspa.1984.0023} {\bibfield  {journal} {\bibinfo  {journal} {Proceedings of the Royal Society of London. A. Mathematical and Physical Sciences}\ }\textbf {\bibinfo {volume} {392}},\ \bibinfo {pages} {45} (\bibinfo {year} {1984})}\BibitemShut {NoStop}%
\bibitem [{\citenamefont {Resta}(1994)}]{Resta1994}%
  \BibitemOpen
  \bibfield  {author} {\bibinfo {author} {\bibfnamefont {R.}~\bibnamefont {Resta}},\ }\bibfield  {title} {\bibinfo {title} {Macroscopic polarization in crystalline dielectrics: the geometric phase approach},\ }\href {https://doi.org/10.1103/RevModPhys.66.899} {\bibfield  {journal} {\bibinfo  {journal} {Reviews of Modern Physics}\ }\textbf {\bibinfo {volume} {66}},\ \bibinfo {pages} {899} (\bibinfo {year} {1994})}\BibitemShut {NoStop}%
\bibitem [{\citenamefont {Xiao}\ \emph {et~al.}(2010)\citenamefont {Xiao}, \citenamefont {Chang},\ and\ \citenamefont {Niu}}]{Xiao2010}%
  \BibitemOpen
  \bibfield  {author} {\bibinfo {author} {\bibfnamefont {D.}~\bibnamefont {Xiao}}, \bibinfo {author} {\bibfnamefont {M.-C.}\ \bibnamefont {Chang}},\ and\ \bibinfo {author} {\bibfnamefont {Q.}~\bibnamefont {Niu}},\ }\bibfield  {title} {\bibinfo {title} {Berry phase effects on electronic properties},\ }\href {https://doi.org/10.1103/RevModPhys.82.1959} {\bibfield  {journal} {\bibinfo  {journal} {Reviews of Modern Physics}\ }\textbf {\bibinfo {volume} {82}},\ \bibinfo {pages} {1959} (\bibinfo {year} {2010})}\BibitemShut {NoStop}%
\bibitem [{\citenamefont {Aharonov}\ and\ \citenamefont {Bohm}(1959)}]{Aharonov1959}%
  \BibitemOpen
  \bibfield  {author} {\bibinfo {author} {\bibfnamefont {Y.}~\bibnamefont {Aharonov}}\ and\ \bibinfo {author} {\bibfnamefont {D.}~\bibnamefont {Bohm}},\ }\bibfield  {title} {\bibinfo {title} {Significance of electromagnetic potentials in the quantum theory},\ }\href {https://doi.org/10.1103/PhysRev.115.485} {\bibfield  {journal} {\bibinfo  {journal} {Physical Review}\ }\textbf {\bibinfo {volume} {115}},\ \bibinfo {pages} {485} (\bibinfo {year} {1959})}\BibitemShut {NoStop}%
\bibitem [{\citenamefont {Peshkin}\ and\ \citenamefont {Tonomura}(1989)}]{Peshkin1989}%
  \BibitemOpen
  \bibfield  {author} {\bibinfo {author} {\bibfnamefont {M.}~\bibnamefont {Peshkin}}\ and\ \bibinfo {author} {\bibfnamefont {A.}~\bibnamefont {Tonomura}},\ }\href {https://doi.org/10.1007/bfb0032076} {\emph {\bibinfo {title} {The Aharonov-Bohm Effect}}},\ \bibinfo {series} {Lecture Notes in Physics}, Vol.\ \bibinfo {volume} {340}\ (\bibinfo  {publisher} {Springer Berlin Heidelberg},\ \bibinfo {year} {1989})\BibitemShut {NoStop}%
\bibitem [{\citenamefont {Weinhold}(1975)}]{weinholdMetricGeometryEquilibrium1975}%
  \BibitemOpen
  \bibfield  {author} {\bibinfo {author} {\bibfnamefont {F.}~\bibnamefont {Weinhold}},\ }\bibfield  {title} {\bibinfo {title} {Metric geometry of equilibrium thermodynamics},\ }\href {https://doi.org/10.1063/1.431689} {\bibfield  {journal} {\bibinfo  {journal} {Journal of Chemical Physics}\ }\textbf {\bibinfo {volume} {63}},\ \bibinfo {pages} {2479} (\bibinfo {year} {1975})}\BibitemShut {NoStop}%
\bibitem [{\citenamefont {Ruppeiner}(1979)}]{ruppeinerThermodynamicsRiemannianGeometric1979b}%
  \BibitemOpen
  \bibfield  {author} {\bibinfo {author} {\bibfnamefont {G.}~\bibnamefont {Ruppeiner}},\ }\bibfield  {title} {\bibinfo {title} {Thermodynamics: A riemannian geometric model},\ }\href {https://doi.org/10.1103/PhysRevA.20.1608} {\bibfield  {journal} {\bibinfo  {journal} {Physical Review A}\ }\textbf {\bibinfo {volume} {20}},\ \bibinfo {pages} {1608} (\bibinfo {year} {1979})}\BibitemShut {NoStop}%
\bibitem [{\citenamefont {Ruppeiner}(1995)}]{georgeRiemannianGeometryThermodynamic1995}%
  \BibitemOpen
  \bibfield  {author} {\bibinfo {author} {\bibfnamefont {G.}~\bibnamefont {Ruppeiner}},\ }\bibfield  {title} {\bibinfo {title} {Riemannian geometry in thermodynamic fluctuation theory},\ }\href {https://doi.org/10.1103/RevModPhys.67.605} {\bibfield  {journal} {\bibinfo  {journal} {Review of Modern Physics}\ }\textbf {\bibinfo {volume} {67}},\ \bibinfo {pages} {605} (\bibinfo {year} {1995})}\BibitemShut {NoStop}%
\bibitem [{\citenamefont {Wang}\ \emph {et~al.}(2026)\citenamefont {Wang}, \citenamefont {Zhao}, \citenamefont {Deng},\ and\ \citenamefont {Ma}}]{wang2026thermodynamic}%
  \BibitemOpen
  \bibfield  {author} {\bibinfo {author} {\bibfnamefont {H.}~\bibnamefont {Wang}}, \bibinfo {author} {\bibfnamefont {L.}~\bibnamefont {Zhao}}, \bibinfo {author} {\bibfnamefont {S.}~\bibnamefont {Deng}},\ and\ \bibinfo {author} {\bibfnamefont {Y.-H.}\ \bibnamefont {Ma}},\ }\bibfield  {title} {\bibinfo {title} {Thermodynamic geometry of relaxation},\ }\bibfield  {journal} {\bibinfo  {journal} {arXiv preprint}\ }\href {https://doi.org/https://doi.org/10.48550/arXiv.2604.15000} {https://doi.org/10.48550/arXiv.2604.15000} (\bibinfo {year} {2026})\BibitemShut {NoStop}%
\bibitem [{\citenamefont {Andresen}\ \emph {et~al.}(1984)\citenamefont {Andresen}, \citenamefont {Salamon},\ and\ \citenamefont {Berry}}]{Andresen1984}%
  \BibitemOpen
  \bibfield  {author} {\bibinfo {author} {\bibfnamefont {B.}~\bibnamefont {Andresen}}, \bibinfo {author} {\bibfnamefont {P.}~\bibnamefont {Salamon}},\ and\ \bibinfo {author} {\bibfnamefont {R.~S.}\ \bibnamefont {Berry}},\ }\bibfield  {title} {\bibinfo {title} {Thermodynamics in finite time},\ }\href {https://doi.org/10.1063/1.2916405} {\bibfield  {journal} {\bibinfo  {journal} {Physics Today}\ }\textbf {\bibinfo {volume} {37}},\ \bibinfo {pages} {62} (\bibinfo {year} {1984})}\BibitemShut {NoStop}%
\bibitem [{\citenamefont {Andresen}\ and\ \citenamefont {Salamon}(2022)}]{Bjarne2022}%
  \BibitemOpen
  \bibfield  {author} {\bibinfo {author} {\bibfnamefont {B.}~\bibnamefont {Andresen}}\ and\ \bibinfo {author} {\bibfnamefont {P.}~\bibnamefont {Salamon}},\ }\bibfield  {title} {\bibinfo {title} {Future perspectives of finite-time thermodynamics},\ }\href {https://doi.org/https://doi.org/10.3390/e24050690} {\bibfield  {journal} {\bibinfo  {journal} {Entropy}\ }\textbf {\bibinfo {volume} {24}},\ \bibinfo {pages} {690} (\bibinfo {year} {2022})}\BibitemShut {NoStop}%
\bibitem [{\citenamefont {Dann}\ \emph {et~al.}(2020)\citenamefont {Dann}, \citenamefont {Kosloff},\ and\ \citenamefont {Salamon}}]{dann2020quantum}%
  \BibitemOpen
  \bibfield  {author} {\bibinfo {author} {\bibfnamefont {R.}~\bibnamefont {Dann}}, \bibinfo {author} {\bibfnamefont {R.}~\bibnamefont {Kosloff}},\ and\ \bibinfo {author} {\bibfnamefont {P.}~\bibnamefont {Salamon}},\ }\bibfield  {title} {\bibinfo {title} {Quantum finite-time thermodynamics: insight from a single qubit engine},\ }\href {https://doi.org/https://doi.org/10.3390/e22111255} {\bibfield  {journal} {\bibinfo  {journal} {Entropy}\ }\textbf {\bibinfo {volume} {22}},\ \bibinfo {pages} {1255} (\bibinfo {year} {2020})}\BibitemShut {NoStop}%
\bibitem [{\citenamefont {Qiu}\ \emph {et~al.}()\citenamefont {Qiu}, \citenamefont {Nomura}, \citenamefont {Zhang} \emph {et~al.}}]{qiu20roadmap}%
  \BibitemOpen
  \bibfield  {author} {\bibinfo {author} {\bibfnamefont {Y.}~\bibnamefont {Qiu}}, \bibinfo {author} {\bibfnamefont {M.}~\bibnamefont {Nomura}}, \bibinfo {author} {\bibfnamefont {Z.}~\bibnamefont {Zhang}}, \emph {et~al.},\ }\bibfield  {title} {\bibinfo {title} {Roadmap on thermodynamics and thermal metamaterials},\ }\href {https://doi.org/10.15302/frontphys.2025.065500} {\bibfield  {journal} {\bibinfo  {journal} {Front. Phys.}\ }\textbf {\bibinfo {volume} {20}},\ \bibinfo {pages} {065500}}\BibitemShut {NoStop}%
\bibitem [{\citenamefont {Salamon}\ and\ \citenamefont {Berry}(1983)}]{Salamon1984}%
  \BibitemOpen
  \bibfield  {author} {\bibinfo {author} {\bibfnamefont {P.}~\bibnamefont {Salamon}}\ and\ \bibinfo {author} {\bibfnamefont {R.~S.}\ \bibnamefont {Berry}},\ }\bibfield  {title} {\bibinfo {title} {Thermodynamic length and dissipated availability},\ }\href {https://doi.org/10.1103/PhysRevLett.51.1127} {\bibfield  {journal} {\bibinfo  {journal} {Physical Review Letters}\ }\textbf {\bibinfo {volume} {51}},\ \bibinfo {pages} {1127} (\bibinfo {year} {1983})}\BibitemShut {NoStop}%
\bibitem [{\citenamefont {Crooks}(2007)}]{Crooks2007}%
  \BibitemOpen
  \bibfield  {author} {\bibinfo {author} {\bibfnamefont {G.~E.}\ \bibnamefont {Crooks}},\ }\bibfield  {title} {\bibinfo {title} {Measuring thermodynamic length},\ }\href {https://doi.org/10.1103/PhysRevLett.99.100602} {\bibfield  {journal} {\bibinfo  {journal} {Physical Review Letters}\ }\textbf {\bibinfo {volume} {99}},\ \bibinfo {pages} {100602} (\bibinfo {year} {2007})}\BibitemShut {NoStop}%
\bibitem [{\citenamefont {Sivak}\ and\ \citenamefont {Crooks}(2012)}]{Sivak2012}%
  \BibitemOpen
  \bibfield  {author} {\bibinfo {author} {\bibfnamefont {D.~A.}\ \bibnamefont {Sivak}}\ and\ \bibinfo {author} {\bibfnamefont {G.~E.}\ \bibnamefont {Crooks}},\ }\bibfield  {title} {\bibinfo {title} {Thermodynamic metrics and optimal paths},\ }\href {https://doi.org/10.1103/PhysRevLett.108.190602} {\bibfield  {journal} {\bibinfo  {journal} {Physical Review Letters}\ }\textbf {\bibinfo {volume} {108}},\ \bibinfo {pages} {190602} (\bibinfo {year} {2012})}\BibitemShut {NoStop}%
\bibitem [{\citenamefont {Shiraishi}\ and\ \citenamefont {Saito}(2019)}]{Naoto2019}%
  \BibitemOpen
  \bibfield  {author} {\bibinfo {author} {\bibfnamefont {N.}~\bibnamefont {Shiraishi}}\ and\ \bibinfo {author} {\bibfnamefont {K.}~\bibnamefont {Saito}},\ }\bibfield  {title} {\bibinfo {title} {Information-theoretical bound of the irreversibility in thermal relaxation processes},\ }\href {https://doi.org/10.1103/PhysRevLett.123.110603} {\bibfield  {journal} {\bibinfo  {journal} {Physical Review Letters}\ }\textbf {\bibinfo {volume} {123}},\ \bibinfo {pages} {110603} (\bibinfo {year} {2019})}\BibitemShut {NoStop}%
\bibitem [{\citenamefont {Shiraishi}(2023)}]{shiraishi2023introduction}%
  \BibitemOpen
  \bibfield  {author} {\bibinfo {author} {\bibfnamefont {N.}~\bibnamefont {Shiraishi}},\ }\bibfield  {title} {\bibinfo {title} {An introduction to stochastic thermodynamics},\ }\bibfield  {journal} {\bibinfo  {journal} {Fundamental Theories of Physics. Springer, Singapore}\ }\href {https://doi.org/https://doi.org/10.1007/978-981-19-8186-9} {https://doi.org/10.1007/978-981-19-8186-9} (\bibinfo {year} {2023})\BibitemShut {NoStop}%
\bibitem [{\citenamefont {Scandi}\ and\ \citenamefont {Perarnau-Llobet}(2019)}]{Scandi2019}%
  \BibitemOpen
  \bibfield  {author} {\bibinfo {author} {\bibfnamefont {M.}~\bibnamefont {Scandi}}\ and\ \bibinfo {author} {\bibfnamefont {M.}~\bibnamefont {Perarnau-Llobet}},\ }\bibfield  {title} {\bibinfo {title} {Thermodynamic length in open quantum systems},\ }\href {https://doi.org/10.22331/q-2019-10-24-197} {\bibfield  {journal} {\bibinfo  {journal} {Quantum}\ }\textbf {\bibinfo {volume} {3}},\ \bibinfo {pages} {197} (\bibinfo {year} {2019})}\BibitemShut {NoStop}%
\bibitem [{\citenamefont {Abiuso}\ \emph {et~al.}(2020)\citenamefont {Abiuso}, \citenamefont {Miller}, \citenamefont {Perarnau-Llobet},\ and\ \citenamefont {Scandi}}]{Abiuso2020e}%
  \BibitemOpen
  \bibfield  {author} {\bibinfo {author} {\bibfnamefont {P.}~\bibnamefont {Abiuso}}, \bibinfo {author} {\bibfnamefont {H.~J.~D.}\ \bibnamefont {Miller}}, \bibinfo {author} {\bibfnamefont {M.}~\bibnamefont {Perarnau-Llobet}},\ and\ \bibinfo {author} {\bibfnamefont {M.}~\bibnamefont {Scandi}},\ }\bibfield  {title} {\bibinfo {title} {Geometric optimisation of quantum thermodynamic processes},\ }\href {https://doi.org/10.3390/e22101076} {\bibfield  {journal} {\bibinfo  {journal} {Entropy}\ }\textbf {\bibinfo {volume} {22}},\ \bibinfo {pages} {1076} (\bibinfo {year} {2020})}\BibitemShut {NoStop}%
\bibitem [{\citenamefont {Anza}\ and\ \citenamefont {Crutchfield}(2022)}]{2022AnzaPRE}%
  \BibitemOpen
  \bibfield  {author} {\bibinfo {author} {\bibfnamefont {F.}~\bibnamefont {Anza}}\ and\ \bibinfo {author} {\bibfnamefont {J.~P.}\ \bibnamefont {Crutchfield}},\ }\bibfield  {title} {\bibinfo {title} {Geometric quantum thermodynamics},\ }\href {https://doi.org/10.1103/PhysRevE.106.054102} {\bibfield  {journal} {\bibinfo  {journal} {Physical Review E}\ }\textbf {\bibinfo {volume} {106}},\ \bibinfo {pages} {054102} (\bibinfo {year} {2022})}\BibitemShut {NoStop}%
\bibitem [{\citenamefont {Ren}\ \emph {et~al.}(2010)\citenamefont {Ren}, \citenamefont {H\"anggi},\ and\ \citenamefont {Li}}]{Ren2010}%
  \BibitemOpen
  \bibfield  {author} {\bibinfo {author} {\bibfnamefont {J.}~\bibnamefont {Ren}}, \bibinfo {author} {\bibfnamefont {P.}~\bibnamefont {H\"anggi}},\ and\ \bibinfo {author} {\bibfnamefont {B.}~\bibnamefont {Li}},\ }\bibfield  {title} {\bibinfo {title} {Berry-phase-induced heat pumping and its impact on the fluctuation theorem},\ }\href {https://doi.org/10.1103/PhysRevLett.104.170601} {\bibfield  {journal} {\bibinfo  {journal} {Physical Review Letters}\ }\textbf {\bibinfo {volume} {104}},\ \bibinfo {pages} {170601} (\bibinfo {year} {2010})}\BibitemShut {NoStop}%
\bibitem [{\citenamefont {Abiuso}\ and\ \citenamefont {Perarnau-Llobet}(2020)}]{Abiuso2020}%
  \BibitemOpen
  \bibfield  {author} {\bibinfo {author} {\bibfnamefont {P.}~\bibnamefont {Abiuso}}\ and\ \bibinfo {author} {\bibfnamefont {M.}~\bibnamefont {Perarnau-Llobet}},\ }\bibfield  {title} {\bibinfo {title} {Geometric properties of adiabatic quantum thermal machines},\ }\href {https://doi.org/10.1103/PhysRevLett.124.110606} {\bibfield  {journal} {\bibinfo  {journal} {Physical Review Letters}\ }\textbf {\bibinfo {volume} {124}},\ \bibinfo {pages} {110606} (\bibinfo {year} {2020})}\BibitemShut {NoStop}%
\bibitem [{\citenamefont {Brandner}\ and\ \citenamefont {Saito}(2020)}]{Brandner2020}%
  \BibitemOpen
  \bibfield  {author} {\bibinfo {author} {\bibfnamefont {K.}~\bibnamefont {Brandner}}\ and\ \bibinfo {author} {\bibfnamefont {K.}~\bibnamefont {Saito}},\ }\bibfield  {title} {\bibinfo {title} {Thermodynamic geometry of microscopic heat engines},\ }\href {https://doi.org/10.1103/PhysRevLett.124.040602} {\bibfield  {journal} {\bibinfo  {journal} {Physical Review Letters}\ }\textbf {\bibinfo {volume} {124}},\ \bibinfo {pages} {040602} (\bibinfo {year} {2020})}\BibitemShut {NoStop}%
\bibitem [{\citenamefont {Bhandari}\ \emph {et~al.}(2020)\citenamefont {Bhandari}, \citenamefont {Alonso}, \citenamefont {Taddei}, \citenamefont {von Oppen}, \citenamefont {Fazio},\ and\ \citenamefont {Arrachea}}]{Bhandari2020}%
  \BibitemOpen
  \bibfield  {author} {\bibinfo {author} {\bibfnamefont {B.}~\bibnamefont {Bhandari}}, \bibinfo {author} {\bibfnamefont {P.~T.}\ \bibnamefont {Alonso}}, \bibinfo {author} {\bibfnamefont {F.}~\bibnamefont {Taddei}}, \bibinfo {author} {\bibfnamefont {F.}~\bibnamefont {von Oppen}}, \bibinfo {author} {\bibfnamefont {R.}~\bibnamefont {Fazio}},\ and\ \bibinfo {author} {\bibfnamefont {L.}~\bibnamefont {Arrachea}},\ }\bibfield  {title} {\bibinfo {title} {Geometric properties of adiabatic quantum thermal machines},\ }\href {https://doi.org/10.1103/PhysRevB.102.155407} {\bibfield  {journal} {\bibinfo  {journal} {Physical Review B}\ }\textbf {\bibinfo {volume} {102}},\ \bibinfo {pages} {155407} (\bibinfo {year} {2020})}\BibitemShut {NoStop}%
\bibitem [{\citenamefont {Berry}\ and\ \citenamefont {Robbins}(1993)}]{Berry1993}%
  \BibitemOpen
  \bibfield  {author} {\bibinfo {author} {\bibfnamefont {M.~V.}\ \bibnamefont {Berry}}\ and\ \bibinfo {author} {\bibfnamefont {J.~M.}\ \bibnamefont {Robbins}},\ }\bibfield  {title} {\bibinfo {title} {Geometric magnetism and deterministic friction},\ }\href {https://doi.org/10.1098/rspa.1993.0127} {\bibfield  {journal} {\bibinfo  {journal} {Proceedings of the Royal Society of London. Series A: Mathematical and Physical Sciences}\ }\textbf {\bibinfo {volume} {442}},\ \bibinfo {pages} {659} (\bibinfo {year} {1993})}\BibitemShut {NoStop}%
\bibitem [{\citenamefont {Campisi}\ \emph {et~al.}(2012)\citenamefont {Campisi}, \citenamefont {Denisov},\ and\ \citenamefont {H\"anggi}}]{Campisi2012}%
  \BibitemOpen
  \bibfield  {author} {\bibinfo {author} {\bibfnamefont {M.}~\bibnamefont {Campisi}}, \bibinfo {author} {\bibfnamefont {S.}~\bibnamefont {Denisov}},\ and\ \bibinfo {author} {\bibfnamefont {P.}~\bibnamefont {H\"anggi}},\ }\bibfield  {title} {\bibinfo {title} {Geometric magnetism in open quantum systems},\ }\href {https://doi.org/10.1103/PhysRevA.86.032114} {\bibfield  {journal} {\bibinfo  {journal} {Physical Review A}\ }\textbf {\bibinfo {volume} {86}},\ \bibinfo {pages} {032114} (\bibinfo {year} {2012})}\BibitemShut {NoStop}%
\bibitem [{\citenamefont {Ma}\ \emph {et~al.}(2022)\citenamefont {Ma}, \citenamefont {Chen}, \citenamefont {Sun},\ and\ \citenamefont {Dong}}]{maMinimalEnergyCost2022}%
  \BibitemOpen
  \bibfield  {author} {\bibinfo {author} {\bibfnamefont {Y.-H.}\ \bibnamefont {Ma}}, \bibinfo {author} {\bibfnamefont {J.-F.}\ \bibnamefont {Chen}}, \bibinfo {author} {\bibfnamefont {C.~P.}\ \bibnamefont {Sun}},\ and\ \bibinfo {author} {\bibfnamefont {H.}~\bibnamefont {Dong}},\ }\bibfield  {title} {\bibinfo {title} {Minimal energy cost to initialize a bit with tolerable error},\ }\href {https://doi.org/10.1103/PhysRevE.106.034112} {\bibfield  {journal} {\bibinfo  {journal} {Physical Review E}\ }\textbf {\bibinfo {volume} {106}},\ \bibinfo {pages} {034112} (\bibinfo {year} {2022})}\BibitemShut {NoStop}%
\bibitem [{\citenamefont {Zurek}(2003)}]{Zurek2003}%
  \BibitemOpen
  \bibfield  {author} {\bibinfo {author} {\bibfnamefont {W.~H.}\ \bibnamefont {Zurek}},\ }\bibfield  {title} {\bibinfo {title} {Decoherence, einselection, and the quantum origins of the classical},\ }\href {https://doi.org/10.1103/RevModPhys.75.715} {\bibfield  {journal} {\bibinfo  {journal} {Reviews of Modern Physics}\ }\textbf {\bibinfo {volume} {75}},\ \bibinfo {pages} {715} (\bibinfo {year} {2003})}\BibitemShut {NoStop}%
\bibitem [{\citenamefont {Alicki}(1979)}]{Alicki1979}%
  \BibitemOpen
  \bibfield  {author} {\bibinfo {author} {\bibfnamefont {R.}~\bibnamefont {Alicki}},\ }\bibfield  {title} {\bibinfo {title} {The quantum open system as a model of the heat engine},\ }\href {https://doi.org/10.1088/0305-4470/12/5/007} {\bibfield  {journal} {\bibinfo  {journal} {Journal of Physics A: Mathematical and General}\ }\textbf {\bibinfo {volume} {12}},\ \bibinfo {pages} {L103} (\bibinfo {year} {1979})}\BibitemShut {NoStop}%
\bibitem [{\citenamefont {Su}\ \emph {et~al.}(2018)\citenamefont {Su}, \citenamefont {Chen}, \citenamefont {Ma}, \citenamefont {Chen},\ and\ \citenamefont {Sun}}]{su2018heat}%
  \BibitemOpen
  \bibfield  {author} {\bibinfo {author} {\bibfnamefont {S.}~\bibnamefont {Su}}, \bibinfo {author} {\bibfnamefont {J.}~\bibnamefont {Chen}}, \bibinfo {author} {\bibfnamefont {Y.}~\bibnamefont {Ma}}, \bibinfo {author} {\bibfnamefont {J.}~\bibnamefont {Chen}},\ and\ \bibinfo {author} {\bibfnamefont {C.}~\bibnamefont {Sun}},\ }\bibfield  {title} {\bibinfo {title} {The heat and work of quantum thermodynamic processes with quantum coherence},\ }\href {https://doi.org/10.1088/1674-1056/27/6/060502} {\bibfield  {journal} {\bibinfo  {journal} {Chinese Physics B}\ }\textbf {\bibinfo {volume} {27}},\ \bibinfo {pages} {060502} (\bibinfo {year} {2018})}\BibitemShut {NoStop}%
\bibitem [{\citenamefont {Seegebrecht}\ and\ \citenamefont {Schilling}(2024)}]{Seegebrecht2024}%
  \BibitemOpen
  \bibfield  {author} {\bibinfo {author} {\bibfnamefont {A.}~\bibnamefont {Seegebrecht}}\ and\ \bibinfo {author} {\bibfnamefont {T.}~\bibnamefont {Schilling}},\ }\bibfield  {title} {\bibinfo {title} {Work, heat and internal energy in open quantum systems: A comparison of four approaches from the autonomous system framework},\ }\href {https://doi.org/10.1007/s10955-024-03251-5} {\bibfield  {journal} {\bibinfo  {journal} {Journal of Statistical Physics}\ }\textbf {\bibinfo {volume} {191}},\ \bibinfo {pages} {34} (\bibinfo {year} {2024})}\BibitemShut {NoStop}%
\bibitem [{\citenamefont {Breuer}\ and\ \citenamefont {Petruccione}(2002)}]{breuer2002theory}%
  \BibitemOpen
  \bibfield  {author} {\bibinfo {author} {\bibfnamefont {H.-P.}\ \bibnamefont {Breuer}}\ and\ \bibinfo {author} {\bibfnamefont {F.}~\bibnamefont {Petruccione}},\ }\href@noop {} {\emph {\bibinfo {title} {The theory of open quantum systems}}}\ (\bibinfo  {publisher} {OUP Oxford},\ \bibinfo {year} {2002})\BibitemShut {NoStop}%
\bibitem [{\citenamefont {Albash}\ \emph {et~al.}(2012)\citenamefont {Albash}, \citenamefont {Boixo}, \citenamefont {Lidar},\ and\ \citenamefont {Zanardi}}]{Albash2012}%
  \BibitemOpen
  \bibfield  {author} {\bibinfo {author} {\bibfnamefont {T.}~\bibnamefont {Albash}}, \bibinfo {author} {\bibfnamefont {S.}~\bibnamefont {Boixo}}, \bibinfo {author} {\bibfnamefont {D.~A.}\ \bibnamefont {Lidar}},\ and\ \bibinfo {author} {\bibfnamefont {P.}~\bibnamefont {Zanardi}},\ }\bibfield  {title} {\bibinfo {title} {Quantum adiabatic markovian master equations},\ }\href {https://doi.org/10.1088/1367-2630/14/12/123016} {\bibfield  {journal} {\bibinfo  {journal} {New Journal of Physics}\ }\textbf {\bibinfo {volume} {14}},\ \bibinfo {pages} {123016} (\bibinfo {year} {2012})}\BibitemShut {NoStop}%
\bibitem [{\citenamefont {Yamaguchi}\ \emph {et~al.}(2017)\citenamefont {Yamaguchi}, \citenamefont {Yuge},\ and\ \citenamefont {Ogawa}}]{Yamaguchi2017}%
  \BibitemOpen
  \bibfield  {author} {\bibinfo {author} {\bibfnamefont {M.}~\bibnamefont {Yamaguchi}}, \bibinfo {author} {\bibfnamefont {T.}~\bibnamefont {Yuge}},\ and\ \bibinfo {author} {\bibfnamefont {T.}~\bibnamefont {Ogawa}},\ }\bibfield  {title} {\bibinfo {title} {Markovian quantum master equation beyond adiabatic regime},\ }\href {https://doi.org/10.1103/PhysRevE.95.012136} {\bibfield  {journal} {\bibinfo  {journal} {Physical Review E}\ }\textbf {\bibinfo {volume} {95}},\ \bibinfo {pages} {012136} (\bibinfo {year} {2017})}\BibitemShut {NoStop}%
\bibitem [{\citenamefont {Sun}(1988)}]{Sun1988}%
  \BibitemOpen
  \bibfield  {author} {\bibinfo {author} {\bibfnamefont {C.-P.}\ \bibnamefont {Sun}},\ }\bibfield  {title} {\bibinfo {title} {Higher-order quantum adiabatic approximation and berry's phase factor},\ }\href {https://doi.org/10.1088/0305-4470/21/7/023} {\bibfield  {journal} {\bibinfo  {journal} {Journal of Physics A: Mathematical and General}\ }\textbf {\bibinfo {volume} {21}},\ \bibinfo {pages} {1595} (\bibinfo {year} {1988})}\BibitemShut {NoStop}%
\bibitem [{\citenamefont {Rigolin}\ \emph {et~al.}(2008)\citenamefont {Rigolin}, \citenamefont {Ortiz},\ and\ \citenamefont {Ponce}}]{Rigolin2008}%
  \BibitemOpen
  \bibfield  {author} {\bibinfo {author} {\bibfnamefont {G.}~\bibnamefont {Rigolin}}, \bibinfo {author} {\bibfnamefont {G.}~\bibnamefont {Ortiz}},\ and\ \bibinfo {author} {\bibfnamefont {V.~H.}\ \bibnamefont {Ponce}},\ }\bibfield  {title} {\bibinfo {title} {Beyond the quantum adiabatic approximation: Adiabatic perturbation theory},\ }\href {https://doi.org/10.1103/PhysRevA.78.052508} {\bibfield  {journal} {\bibinfo  {journal} {Physical Review A}\ }\textbf {\bibinfo {volume} {78}},\ \bibinfo {pages} {052508} (\bibinfo {year} {2008})}\BibitemShut {NoStop}%
\bibitem [{\citenamefont {Ma}\ \emph {et~al.}(2018)\citenamefont {Ma}, \citenamefont {Xu}, \citenamefont {Dong},\ and\ \citenamefont {Sun}}]{maUniversalConstraintEfficiency2018}%
  \BibitemOpen
  \bibfield  {author} {\bibinfo {author} {\bibfnamefont {Y.-H.}\ \bibnamefont {Ma}}, \bibinfo {author} {\bibfnamefont {D.}~\bibnamefont {Xu}}, \bibinfo {author} {\bibfnamefont {H.}~\bibnamefont {Dong}},\ and\ \bibinfo {author} {\bibfnamefont {C.-P.}\ \bibnamefont {Sun}},\ }\bibfield  {title} {\bibinfo {title} {Universal constraint for efficiency and power of a low-dissipation heat engine},\ }\href {https://doi.org/10.1103/PhysRevE.98.042112} {\bibfield  {journal} {\bibinfo  {journal} {Physical Review E}\ }\textbf {\bibinfo {volume} {98}},\ \bibinfo {pages} {042112} (\bibinfo {year} {2018})}\BibitemShut {NoStop}%
\bibitem [{\citenamefont {Ma}\ \emph {et~al.}(2020)\citenamefont {Ma}, \citenamefont {Zhai}, \citenamefont {Chen}, \citenamefont {Sun},\ and\ \citenamefont {Dong}}]{ma2020experimental}%
  \BibitemOpen
  \bibfield  {author} {\bibinfo {author} {\bibfnamefont {Y.-H.}\ \bibnamefont {Ma}}, \bibinfo {author} {\bibfnamefont {R.-X.}\ \bibnamefont {Zhai}}, \bibinfo {author} {\bibfnamefont {J.}~\bibnamefont {Chen}}, \bibinfo {author} {\bibfnamefont {C.}~\bibnamefont {Sun}},\ and\ \bibinfo {author} {\bibfnamefont {H.}~\bibnamefont {Dong}},\ }\bibfield  {title} {\bibinfo {title} {Experimental test of the 1/$\tau$-scaling entropy generation in finite-time thermodynamics},\ }\href {https://doi.org/10.1103/PhysRevLett.125.210601} {\bibfield  {journal} {\bibinfo  {journal} {Physical Review Letters}\ }\textbf {\bibinfo {volume} {125}},\ \bibinfo {pages} {210601} (\bibinfo {year} {2020})}\BibitemShut {NoStop}%
\bibitem [{\citenamefont {Yuan}\ \emph {et~al.}(2022)\citenamefont {Yuan}, \citenamefont {Ma},\ and\ \citenamefont {Sun}}]{yuanOptimizingThermodynamicCycles2022}%
  \BibitemOpen
  \bibfield  {author} {\bibinfo {author} {\bibfnamefont {H.}~\bibnamefont {Yuan}}, \bibinfo {author} {\bibfnamefont {Y.-H.}\ \bibnamefont {Ma}},\ and\ \bibinfo {author} {\bibfnamefont {C.~P.}\ \bibnamefont {Sun}},\ }\bibfield  {title} {\bibinfo {title} {Optimizing thermodynamic cycles with two finite-sized reservoirs},\ }\href {https://doi.org/10.1103/PhysRevE.105.L022101} {\bibfield  {journal} {\bibinfo  {journal} {Physical Review E}\ }\textbf {\bibinfo {volume} {105}},\ \bibinfo {pages} {L022101} (\bibinfo {year} {2022})}\BibitemShut {NoStop}%
\bibitem [{\citenamefont {Zhao}\ \emph {et~al.}(2024)\citenamefont {Zhao}, \citenamefont {Tu},\ and\ \citenamefont {Ma}}]{zhaoEngineeringRatchetbasedParticle2024}%
  \BibitemOpen
  \bibfield  {author} {\bibinfo {author} {\bibfnamefont {X.-H.}\ \bibnamefont {Zhao}}, \bibinfo {author} {\bibfnamefont {Z.~C.}\ \bibnamefont {Tu}},\ and\ \bibinfo {author} {\bibfnamefont {Y.-H.}\ \bibnamefont {Ma}},\ }\bibfield  {title} {\bibinfo {title} {Engineering ratchet-based particle separation via extended shortcuts to isothermality},\ }\href {https://doi.org/10.1103/PhysRevE.110.034105} {\bibfield  {journal} {\bibinfo  {journal} {Physical Review E}\ }\textbf {\bibinfo {volume} {110}},\ \bibinfo {pages} {034105} (\bibinfo {year} {2024})}\BibitemShut {NoStop}%
\bibitem [{\citenamefont {Li}\ \emph {et~al.}(2022)\citenamefont {Li}, \citenamefont {Chen}, \citenamefont {Sun},\ and\ \citenamefont {Dong}}]{liGeodesicPathMinimal2022}%
  \BibitemOpen
  \bibfield  {author} {\bibinfo {author} {\bibfnamefont {G.}~\bibnamefont {Li}}, \bibinfo {author} {\bibfnamefont {J.-F.}\ \bibnamefont {Chen}}, \bibinfo {author} {\bibfnamefont {C.~P.}\ \bibnamefont {Sun}},\ and\ \bibinfo {author} {\bibfnamefont {H.}~\bibnamefont {Dong}},\ }\bibfield  {title} {\bibinfo {title} {Geodesic {{Path}} for the {{Minimal Energy Cost}} in {{Shortcuts}} to {{Isothermality}}},\ }\href {https://doi.org/10.1103/PhysRevLett.128.230603} {\bibfield  {journal} {\bibinfo  {journal} {Physical Review Letters}\ }\textbf {\bibinfo {volume} {128}},\ \bibinfo {pages} {230603} (\bibinfo {year} {2022})}\BibitemShut {NoStop}%
\bibitem [{\citenamefont {Chen}\ \emph {et~al.}(2019)\citenamefont {Chen}, \citenamefont {Sun},\ and\ \citenamefont {Dong}}]{Chen2019}%
  \BibitemOpen
  \bibfield  {author} {\bibinfo {author} {\bibfnamefont {J.-F.}\ \bibnamefont {Chen}}, \bibinfo {author} {\bibfnamefont {C.-P.}\ \bibnamefont {Sun}},\ and\ \bibinfo {author} {\bibfnamefont {H.}~\bibnamefont {Dong}},\ }\bibfield  {title} {\bibinfo {title} {Boosting the performance of quantum otto heat engines},\ }\href {https://doi.org/10.1103/PhysRevE.100.032144} {\bibfield  {journal} {\bibinfo  {journal} {Physical Review E}\ }\textbf {\bibinfo {volume} {100}},\ \bibinfo {pages} {032144} (\bibinfo {year} {2019})}\BibitemShut {NoStop}%
\bibitem [{\citenamefont {Fei}\ \emph {et~al.}(2022)\citenamefont {Fei}, \citenamefont {Chen},\ and\ \citenamefont {Ma}}]{fei2022efficiency}%
  \BibitemOpen
  \bibfield  {author} {\bibinfo {author} {\bibfnamefont {Z.}~\bibnamefont {Fei}}, \bibinfo {author} {\bibfnamefont {J.-F.}\ \bibnamefont {Chen}},\ and\ \bibinfo {author} {\bibfnamefont {Y.-H.}\ \bibnamefont {Ma}},\ }\bibfield  {title} {\bibinfo {title} {Efficiency statistics of a quantum otto cycle},\ }\href {https://doi.org/10.1103/PhysRevA.105.022609} {\bibfield  {journal} {\bibinfo  {journal} {Physical Review A}\ }\textbf {\bibinfo {volume} {105}},\ \bibinfo {pages} {022609} (\bibinfo {year} {2022})}\BibitemShut {NoStop}%
\bibitem [{\citenamefont {Liu}(2020)}]{liu2020fluctuation}%
  \BibitemOpen
  \bibfield  {author} {\bibinfo {author} {\bibfnamefont {F.}~\bibnamefont {Liu}},\ }\bibfield  {title} {\bibinfo {title} {A fluctuation theorem for floquet quantum master equations},\ }\href {https://doi.org/10.1088/1572-9494/ab95fc} {\bibfield  {journal} {\bibinfo  {journal} {Communications in Theoretical Physics}\ }\textbf {\bibinfo {volume} {72}},\ \bibinfo {pages} {095601} (\bibinfo {year} {2020})}\BibitemShut {NoStop}%
\bibitem [{\citenamefont {Batalh\~ao}\ \emph {et~al.}(2014)\citenamefont {Batalh\~ao}, \citenamefont {Souza}, \citenamefont {Mazzola}, \citenamefont {Auccaise}, \citenamefont {Sarthour}, \citenamefont {Oliveira}, \citenamefont {Goold}, \citenamefont {De~Chiara}, \citenamefont {Paternostro},\ and\ \citenamefont {Serra}}]{2014NMR}%
  \BibitemOpen
  \bibfield  {author} {\bibinfo {author} {\bibfnamefont {T.~B.}\ \bibnamefont {Batalh\~ao}}, \bibinfo {author} {\bibfnamefont {A.~M.}\ \bibnamefont {Souza}}, \bibinfo {author} {\bibfnamefont {L.}~\bibnamefont {Mazzola}}, \bibinfo {author} {\bibfnamefont {R.}~\bibnamefont {Auccaise}}, \bibinfo {author} {\bibfnamefont {R.~S.}\ \bibnamefont {Sarthour}}, \bibinfo {author} {\bibfnamefont {I.~S.}\ \bibnamefont {Oliveira}}, \bibinfo {author} {\bibfnamefont {J.}~\bibnamefont {Goold}}, \bibinfo {author} {\bibfnamefont {G.}~\bibnamefont {De~Chiara}}, \bibinfo {author} {\bibfnamefont {M.}~\bibnamefont {Paternostro}},\ and\ \bibinfo {author} {\bibfnamefont {R.~M.}\ \bibnamefont {Serra}},\ }\bibfield  {title} {\bibinfo {title} {Experimental reconstruction of work distribution and study of fluctuation relations in a closed quantum system},\ }\href {https://doi.org/10.1103/PhysRevLett.113.140601} {\bibfield  {journal} {\bibinfo  {journal} {Physical Review Letters}\ }\textbf {\bibinfo {volume} {113}},\ \bibinfo {pages}
  {140601} (\bibinfo {year} {2014})}\BibitemShut {NoStop}%
\bibitem [{\citenamefont {Vieira}\ \emph {et~al.}(2023)\citenamefont {Vieira}, \citenamefont {{de Oliveira}}, \citenamefont {Santos}, \citenamefont {Dieguez},\ and\ \citenamefont {Serra}}]{2023NMR}%
  \BibitemOpen
  \bibfield  {author} {\bibinfo {author} {\bibfnamefont {C.}~\bibnamefont {Vieira}}, \bibinfo {author} {\bibfnamefont {J.}~\bibnamefont {{de Oliveira}}}, \bibinfo {author} {\bibfnamefont {J.}~\bibnamefont {Santos}}, \bibinfo {author} {\bibfnamefont {P.}~\bibnamefont {Dieguez}},\ and\ \bibinfo {author} {\bibfnamefont {R.}~\bibnamefont {Serra}},\ }\bibfield  {title} {\bibinfo {title} {Exploring quantum thermodynamics with nmr},\ }\href {https://doi.org/https://doi.org/10.1016/j.jmro.2023.100105} {\bibfield  {journal} {\bibinfo  {journal} {Journal of Magnetic Resonance Open}\ }\textbf {\bibinfo {volume} {16-17}},\ \bibinfo {pages} {100105} (\bibinfo {year} {2023})}\BibitemShut {NoStop}%
\bibitem [{\citenamefont {Ulyanov}\ and\ \citenamefont {Zaslavskii}(1992)}]{ULYANOV1992179}%
  \BibitemOpen
  \bibfield  {author} {\bibinfo {author} {\bibfnamefont {V.}~\bibnamefont {Ulyanov}}\ and\ \bibinfo {author} {\bibfnamefont {O.}~\bibnamefont {Zaslavskii}},\ }\bibfield  {title} {\bibinfo {title} {New methods in the theory of quantum spin systems},\ }\href {https://doi.org/https://doi.org/10.1016/0370-1573(92)90158-V} {\bibfield  {journal} {\bibinfo  {journal} {Physics Reports}\ }\textbf {\bibinfo {volume} {216}},\ \bibinfo {pages} {179} (\bibinfo {year} {1992})}\BibitemShut {NoStop}%
\bibitem [{\citenamefont {Ma}\ \emph {et~al.}(2017)\citenamefont {Ma}, \citenamefont {Su},\ and\ \citenamefont {Sun}}]{ma2017quantum}%
  \BibitemOpen
  \bibfield  {author} {\bibinfo {author} {\bibfnamefont {Y.-H.}\ \bibnamefont {Ma}}, \bibinfo {author} {\bibfnamefont {S.-H.}\ \bibnamefont {Su}},\ and\ \bibinfo {author} {\bibfnamefont {C.-P.}\ \bibnamefont {Sun}},\ }\bibfield  {title} {\bibinfo {title} {Quantum thermodynamic cycle with quantum phase transition},\ }\href {https://doi.org/https://doi.org/10.1103/PhysRevE.96.022143} {\bibfield  {journal} {\bibinfo  {journal} {Physical Review E}\ }\textbf {\bibinfo {volume} {96}},\ \bibinfo {pages} {022143} (\bibinfo {year} {2017})}\BibitemShut {NoStop}%
\bibitem [{\citenamefont {Ma}\ and\ \citenamefont {Sun}(2017)}]{ma2017quantum1}%
  \BibitemOpen
  \bibfield  {author} {\bibinfo {author} {\bibfnamefont {Y.-H.}\ \bibnamefont {Ma}}\ and\ \bibinfo {author} {\bibfnamefont {C.-P.}\ \bibnamefont {Sun}},\ }\bibfield  {title} {\bibinfo {title} {Quantum sensing of rotation velocity based on transverse field ising model},\ }\href {https://doi.org/https://doi.org/10.1140/epjd/e2017-80247-x} {\bibfield  {journal} {\bibinfo  {journal} {The European Physical Journal D}\ }\textbf {\bibinfo {volume} {71}},\ \bibinfo {pages} {249} (\bibinfo {year} {2017})}\BibitemShut {NoStop}%
\bibitem [{\citenamefont {Yi}\ \emph {et~al.}(2026)\citenamefont {Yi}, \citenamefont {Qiao}, \citenamefont {Yue},\ and\ \citenamefont {Sun}}]{yi2026third}%
  \BibitemOpen
  \bibfield  {author} {\bibinfo {author} {\bibfnamefont {M.-M.}\ \bibnamefont {Yi}}, \bibinfo {author} {\bibfnamefont {G.-J.}\ \bibnamefont {Qiao}}, \bibinfo {author} {\bibfnamefont {X.}~\bibnamefont {Yue}},\ and\ \bibinfo {author} {\bibfnamefont {C.}~\bibnamefont {Sun}},\ }\bibfield  {title} {\bibinfo {title} {Third quantization for order parameters (ii): Local field quantization in superconducting quantum circuits},\ }\bibfield  {journal} {\bibinfo  {journal} {arXiv preprint}\ }\href {https://doi.org/https://doi.org/10.48550/arXiv.2604.24092} {https://doi.org/10.48550/arXiv.2604.24092} (\bibinfo {year} {2026})\BibitemShut {NoStop}%
\end{thebibliography}%

\clearpage
\onecolumngrid{}

\begin{center}
\textbf{\Large{End Matter}}
\end{center}

\twocolumngrid{}

\textit{Appendix A: Derivation of \rm{[Eq.~\eqref{eq:W1_closed}]}.---}
Integrating the off-diagonal master equation~\eqref{eq:offdiag} with the initial condition $\rho_{mn}(0)=0$ gives the exact coherence as a sum of population-driven and multilevel-feedback parts, $\rho_{mn}(t) = \rho_{mn}^{(\rm p)}(t) + \rho_{mn}^{(\rm f)}(t)$, where
\begin{equation}
  \begin{split}
\rho_{mn}^{(\rm p)}(t)&\equiv \mathrm{i}\int_0^t \, U_{mn}(t,t')\, A_{mn}(t')[\rho_{nn}(t')-\rho_{mm}(t')]\mathrm{d}t',\\
\rho_{mn}^{(\rm f)}(t)&\equiv  \mathrm{i}\int_0^t \, U_{mn}(t,t')\,\mathcal{F}_{mn}[\rho](t')\mathrm{d}t'.
  \end{split}
\end{equation}
Here, $U_{mn}(t,t') = \exp\!\big[-\int_{t'}^t z_{mn}(\tau)\,\mathrm{d}\tau\big]$ and $\mathcal{F}_{mn}[\rho] \equiv \sum_{k\neq m,n}(A_{mk}\rho_{kn}-\rho_{mk}A_{kn})$.

To leading order in the adiabatic parameter $\epsilon$, we use two key simplifications: (i) the population difference is replaced by its initial value, $\rho_{mm}(t)-\rho_{nn}(t) \simeq \Delta_{mn}$, since detailed balance ensures nonadiabatic population corrections are $\mathcal{O}(\epsilon^2)$; (ii) the feedback term is evaluated on the population-driven coherence, $\mathcal{F}_{mn}[\rho](t') = \mathcal{F}_{mn}[\rho^{(\rm p)}](t') + \mathcal{O}(\epsilon^3)$. The coherences then reduce to
\begin{align}
    \rho_{mn}^{(\rm p)}(t) &= -\mathrm{i}\Delta_{mn} c_{mn}(t) + \mathcal{O}(\epsilon^3), \label{eq:rho_p}\\
    \rho_{mn}^{(\rm f)}(t) &= \mathrm{i}\int_0^t \, U_{mn}(t,t')\,\mathcal{F}_{mn}[\rho^{(\rm p)}](t')\mathrm{d}t' + \mathcal{O}(\epsilon^3), \label{eq:rho_f}
\end{align}
where we introduced the dissipative transition amplitude
\begin{equation}\label{eq:c_def}
    c_{mn}(t) \equiv \int_0^t \, U_{mn}(t,t')A_{mn}(t')\mathrm{d}t'.
\end{equation}
Note that $c_{mn}(t)$ reduces to the standard non-adiabatic transition amplitude when $\gamma_{mn}=0$.

Substituting the coherence decomposition into Eq.~\eqref{eq:Woff_def} yields two distinct contributions, $W(t) = W^{(\rm p)}(t) + W^{(\rm f)}(t)$, defined as
\begin{align}
W^{(\rm p)}(t) &\equiv -\hbar\sum_{m\neq n} \omega_{mn} \int_0^t \operatorname{Im}[A_{nm}(t')\rho_{mn}^{(\rm p)}(t')]\mathrm{d}t', \label{eq:Wp_def} \\
W^{(\rm f)}(t) &\equiv -\hbar\sum_{m\neq n} \omega_{mn} \int_0^t \operatorname{Im}[A_{nm}(t')\rho_{mn}^{(\rm f)}(t')]\mathrm{d}t'. \label{eq:Wf_def}
\end{align}
Inserting Eq.~\eqref{eq:rho_p} into Eq.~\eqref{eq:Wp_def} gives the  population-driven contribution  in terms of the dissipative amplitude,
\begin{equation}
W^{(\rm p)}(t) = -\hbar\sum_{m\neq n} \omega_{mn}\Delta_{mn} \int_0^t \operatorname{Re}[A_{nm}(t')c_{mn}(t')]\mathrm{d}t'.
\label{eq:Wp_integral}
\end{equation}

To evaluate the memory integral, we use the integration by parts because $|z_{mn}|\gg 1/T$. Differentiating Eq.~\eqref{eq:c_def} yields the exact relation $\dot{c}_{mn} = A_{mn} - z_{mn}c_{mn}$. Solving for $c_{mn}$ and inserting into a generic integral gives the identity
\begin{equation}
\begin{split}
\int_0^t  g(t')c_{mn}(t')\mathrm{d}t' =& \int_0^t \frac{A_{mn}(t')g(t')}{z_{mn}(t')}\mathrm{d}t' - \left.\frac{g(t')\,c_{mn}(t')}{z_{mn}(t')}\right|_0^t \\
&+ \int_0^t c_{mn}(t') \frac{\mathrm{d}}{\mathrm{d}t'}\!\left[\frac{g(t')}{z_{mn}(t')}\right] \mathrm{d}t'.
\label{eq:by_parts}
\end{split}
\end{equation}
Applying Eq.~\eqref{eq:by_parts} to Eq.~\eqref{eq:Wp_integral} with $g=A_{nm}$ and approximating the remaining $c_{mn}$ by its leading-order form $c_{mn} \simeq A_{mn}/z_{mn}$, the work $W^{(\rm p)}(T)$ over a closed loop $\mathbf{R}(T)=\mathbf{R}(0)$ reduces to boundary contribution at $t'=T$ (the $t'=0$ contribution vanishes because  $c_{mn}(0)=0$), together with continuous integrals over the loop. 

The boundary term  naturally splits into two parts: the first is proportional to $\operatorname{Re}[|A_{mn}|^2/z_{mn}^2]$ and yields the contribution $\cos\varphi_{mn}Q_{mn}$, while the second retains the memory factor $\mathrm{e}^{-\int_0^T z_{mn}\mathrm{d}\tau}$ and produces the damped oscillatory envelope $-Q_{mn}\mathrm{e}^{-\gamma_{mn}T}\cos\Theta_{mn}$. 
The continuous integrals furnish the remaining pieces: the direct integral gives $\gamma_{mn}\!\int_0^T Q_{mn}(t)\mathrm{d}t$ together with the part of $\int_0^T \Lambda_{mn}(t)\mathrm{d}t$ that involves the instantaneous phases $A_{mm}-A_{nn}$, while the derivative term contributes the $\partial_t\arg A_{mn}$ component within $\Lambda_{mn}$. Synthesizing these boundary and continuous integral terms recovers Eq.~\eqref{eq:W1_closed}.

To evaluate $W^{(\rm f)}(t)$, we start from $\rho_{mn}^{(\rm f)}(t)$ in Eq.~\eqref{eq:rho_f}. Using $\partial_{t'}U_{mn}(t,t') = U_{mn}(t,t')z_{mn}(t')$ and integrating by parts gives
\begin{equation}
\begin{split}
\rho_{mn}^{(\rm f)}(t) =& \; \mathrm{i} \left. \frac{\mathcal{F}_{mn}[\rho^{(\rm p)}](t')}{z_{mn}(t')} U_{mn}(t,t') \right |_0^t \\
&- \mathrm{i}\int_0^t U_{mn}(t,t') \frac{\mathrm{d}}{\mathrm{d}t'}\!\left(\frac{\mathcal{F}_{mn}}{z_{mn}}\right) \mathrm{d}t'.
\end{split}
\end{equation}
Since $\mathcal{F}_{mn}/z_{mn} \sim \mathcal{O}(\epsilon^2)$, the boundary term is $\mathcal{O}(\epsilon^2)$. The remaining integral is $\mathcal{O}(\epsilon^3)$ due to the derivative and thus negligible. Keeping only the $t'=t$ boundary term (the $t'=0$ contribution vanishes because $\mathcal{F}_{mn}[\rho^{(\rm p)}](0)=0$) yields the result
\begin{equation}
\rho_{mn}^{(\rm f)}(t) = \frac{\mathrm{i}\mathcal{F}_{mn}[\rho^{(\rm p)}](t)}{z_{mn}(t)} + \mathcal{O}(\epsilon^3).
\label{eq:rho_f_instant}
\end{equation}
Substituting Eq.~\eqref{eq:rho_f_instant} into Eq.~\eqref{eq:Wf_def} and approximating $c_{mn}$ by its leading-order form $c_{mn} \simeq A_{mn}/z_{mn}$, the multilevel-feedback contribution becomes
\begin{equation}
\begin{split}
W^{(\rm f)}=& \; 2\hbar\sum_{m\neq n}\sum_{k\neq m,n} \omega_{mn}\Delta_{mk} \\
&\times \operatorname{Im}\Bigg[ \int_0^T \frac{A_{mk}(t)A_{kn}(t)A_{nm}(t)}{z_{mk}(t)z_{mn}(t)}\mathrm{d}t \Bigg].
\label{eq:W2_after_parts}
\end{split}
\end{equation}
This contribution requires at least three distinct levels linked by non-vanishing cyclic products $A_{mk}A_{kn}A_{nm}$.

\textit{Appendix B: Exact Solution of the Driven Two-Level System.---} Here, we verify the exactness of our general framework by directly solving the Bloch equations for the TLS. Working explicitly in the instantaneous adiabatic basis $\{|+\rangle, |-\rangle\}$, the density matrix and the time derivative of the Hamiltonian \eqref{eq:hamiltonian_tls} read
\begin{equation}
 \hat{\rho}(t) = \frac{1}{2} \begin{pmatrix} 1 + z(t) & x(t) - \mathrm{i} y(t) \\ x(t) + \mathrm{i} y(t) & 1 - z(t) \end{pmatrix}, 
\end{equation}
and
\begin{equation}
 \partial_t\hat{H}(t) =-\frac{\hbar\omega}{2} \begin{pmatrix} 0 & \mathrm{i}\Omega\sin\theta \\ -\mathrm{i}\Omega\sin\theta & 0 \end{pmatrix}, 
\end{equation}
respectively. In this representation, $z(t)$ denotes the instantaneous population difference, while the off-diagonal elements $x(t)$ and $y(t)$ capture the quantum coherences. Evaluating the average work over a closed loop~\eqref{eq:Woff_def} reduces to 
\begin{equation}
W=\int_0^T \mathrm{Tr}\left[ \hat{\rho}(t)\partial_t\hat{H}(t)\right]\mathrm{d}t=\frac{\hbar \omega \Omega \sin\theta }{2}\int_0^T y(t)\, \mathrm{d}t.
\label{eq:work_integral_app}
\end{equation}

The system relaxes toward the instantaneous thermal state with population difference $\Delta = \tanh(\beta\hbar\omega/2)$. 
Let $\gamma'$ and $\gamma$ denote the longitudinal and transverse relaxation rates, respectively. The adiabatic-frame Bloch equations are~\cite{Scandi2019}
\begin{gather}
    \begin{split}
    \dot{x}(t) &= -\tilde{\omega} y(t) - \gamma x(t),  \\
    \dot{y}(t) &= \tilde{\omega} x(t) + \Omega\sin\theta\,z(t) - \gamma y(t),  \\
    \dot{z}(t) &= -\Omega\sin\theta\,y(t) - \gamma'[z(t)-\Delta]\label{eq:bloch}
    \end{split}
\end{gather}
with the initial condition $x(0)=y(0)=0$, $z(0)=\Delta$, and $\tilde{\omega} = \omega - \Omega\cos\theta$. Setting $z(t) \simeq \Delta$, Eq.~\eqref{eq:bloch} reduce to a linear system for $x$ and $y$, with solution
\begin{equation}
y(t) = \frac{\Delta\Omega\sin\theta}{\tilde{\omega}^2 + \gamma^2} \Big[ \gamma - \gamma \mathrm{e}^{-\gamma t}\cos(\tilde{\omega} t) + \tilde{\omega} \mathrm{e}^{-\gamma t}\sin(\tilde{\omega} t) \Big].
\label{eq:y_exact_solution}
\end{equation}
Substituting Eq.~\eqref{eq:y_exact_solution} back into the equation for  $z(t)$ in Eq.~\eqref{eq:bloch} yields
\begin{gather}
    \begin{split}
&z(t)= \Delta - \frac{\Delta\,\Omega^2\sin^2\theta}{\tilde{\omega}^2+\gamma^2}\bigg\{\frac{\gamma}{\gamma'} 
+ \frac{(\gamma' - \gamma)(\gamma^2 + \tilde{\omega}^2)}{\gamma'\bigl[(\gamma' - \gamma)^2 + \tilde{\omega}^2\bigr]} \mathrm{e}^{-\gamma' t} \\
&- \frac{\bigl[\gamma(\gamma' - \gamma) + \tilde{\omega}^2\bigr]\cos(\tilde{\omega} t) - \tilde{\omega}(\gamma' - 2\gamma)\sin(\tilde{\omega} t)}{(\gamma' - \gamma)^2 + \tilde{\omega}^2} \mathrm{e}^{-\gamma t}\bigg\}.
\label{eq:z_solution}
    \end{split}
\end{gather}
The deviation $z(t)-\Delta$ contains only terms proportional to $\Omega^2/\tilde{\omega}^2 \sim \epsilon^2$, confirming the validity of the adiabatic population approximation $z(t) \simeq \Delta$.

Substituting Eq.~\eqref{eq:y_exact_solution} into Eq.~\eqref{eq:work_integral_app}, we obtain the complete non-adiabatic work
\begin{equation}
\begin{split}
W = &\frac{\hbar \omega \Omega^2 \sin^2\theta \Delta}{2(\tilde{\omega}^2 + \gamma^2)^2} \big\{ \gamma(\tilde{\omega}^2 + \gamma^2) T + (\tilde{\omega}^2-\gamma^2) \\
&- \mathrm{e}^{-\gamma T}[ (\tilde{\omega}^2-\gamma^2)\cos(\tilde{\omega} T)+2\gamma\tilde{\omega}\sin(\tilde{\omega} T) ] \big\}.
\end{split}
\label{eq:W_exact_full}
\end{equation}
In the dissipative regime ($\gamma\gg T^{-1}$), $\mathrm{e}^{-\gamma T}\rightarrow0$, $\tilde{\omega}\approx\omega$, and the average work reduces to
\begin{equation}
W = \frac{2\pi^2\Delta\sin^2\theta\,\hbar\omega\gamma}{(\omega^2 + \gamma^2)T} + \mathcal{O}(\epsilon^{2}),
\label{eq:W_avg_limit}
\end{equation}
which, using Eq.~\eqref{LTSL}, becomes $W = \hbar\mathcal{L}^2/T$ up to the leading order. The saturation of the bound in Eq.~\eqref{eq:wpmg} follows from the constant thermodynamic speed maintained by uniform precession.

Subtracting the average work of clockwise ($\Omega$) and counterclockwise ($-\Omega$) protocols leaves only chiral contributions at $\mathcal{O}(T^{-2})$. Extracting these from Eq.~\eqref{eq:W_exact_full}, we begin with the  term $\propto\gamma T$, where expanding the denominator with respect to the driving frequency $\Omega$ yields
\begin{equation}
\frac{\gamma T}{\tilde{\omega}_{\pm}^2 + \gamma^2} \approx \frac{\gamma T}{\omega^2 + \gamma^2} \mp \frac{2\omega\gamma\,\Phi}{(\omega^2 + \gamma^2)^2},
\end{equation}
where $\tilde{\omega}_{\pm} = \omega \pm \Omega\cos\theta $, and $\Phi = 2\pi\cos\theta$ is the Berry phase difference. 
This expansion directly contributes to the chiral work difference.
For the term $\propto\mathrm{e}^{-\gamma T}$, the prefactor $\Omega^2 \sim \mathcal{O}(T^{-2})$ permits setting $\tilde{\omega}_{\pm} \approx \omega_0$ throughout, while the phase shift in the oscillatory functions are expanded as $\cos(\omega T \mp \Phi) = \cos(\omega T)\cos\Phi \pm \sin(\omega T)\sin\Phi$; the subtraction $\Delta W = W_{\rm cw} - W_{\rm cc}$ then isolates the $\sin\Phi$ components. 
Combining these chiral contributions yields
\begin{gather}
    \begin{split}
\Delta W_{\rm TLS} =&\frac{2\hbar\mathcal{L}^2}{\gamma(\omega^2+\gamma^2) T^2} \big\{\mathrm{e}^{-\gamma T}[2\omega\gamma\cos(\omega T)\\
&-(\omega^2-\gamma^2)\sin(\omega T)]\sin\Phi + 2\omega\gamma\Phi\big\},
\label{eq:W_diff_verify}
    \end{split}
\end{gather}
Noting $\varphi=\arctan[2\omega\gamma/(\omega^2-\gamma^2)]$,
the exact identity between Eq.~\eqref{eq:W_diff_verify} and Eq.~\eqref{eq:Delta_W_TLS_exact} provides a stringent validation of the DAPE framework.

\end{document}